\documentclass[12pt]{revtex4}
\usepackage{amssymb,epsf}
\begin{document}

\title{Thermodynamics of $d$-dimensional charged rotating black brane\\
and AdS/CFT correspondence}
\author{M. H. Dehghani$^{1,2}$}\email{dehghani@physics.susc.ac.ir} \author{A. Khodam-Mohammadi$^{1}$}
\address{${1}$. Physics Department and Biruni Observatory, College of Sciences, Shiraz
University, Shiraz 71454, Iran\\
        ${2}$. Institute for Studies in Theoretical Physics and Mathematics (IPM), P.O. Box 19395-5531, Tehran, Iran}
\begin{abstract}
We compute the Euclidean actions of a $d$-dimensional charged
rotating black brane both in the canonical and the grand-canonical
ensemble through the use of the counterterms renormalization
method, and show that the logarithmic divergencies associated with
the Weyl anomalies and matter field vanish. We obtain a Smarr-type
formula for the mass as a function of the entropy, the angular
momenta, and the electric charge, and show that these quantities
satisfy the first law of thermodynamics. Using the conserved
quantities and the Euclidean actions, we calculate the
thermodynamics potentials of the system in terms of the
temperature, angular velocities, and electric potential both in
the canonical and grand-canonical ensembles. We also perform a
stability analysis in these two ensembles, and show that the
system is thermally stable. This is commensurate with the fact
that there is no Hawking-Page phase transition for a black object
with zero curvature horizon. Finally, we obtain the logarithmic
correction of the entropy due to the thermal fluctuation around
the equilibrium.
\end{abstract}

\maketitle

%\pacs{PACS number(s): 04.70.Dy, 04.62.+v, 04.60.-m}

\section{Introduction}

The thermodynamics of asymptotically anti de-Sitter (AAdS) black
holes has attracted a great deal of attention in recent years.
First this is due to the fact that it produces an aggregate of
ideas from thermodynamics, quantum field theory and general
relativity, and second because of its role in the AdS/CFT duality
\cite{Mal}. According to this duality the low energy limit of
string theory in AAdS (times some compact manifold) is equivalent
to quantum field theory living on the boundary of AAdS. This
equivalence between the two formulations means that, at least in
principle, one can obtain complete information on one side of the
duality by performing a computation on the other side. For
example, one can gain some insight into the thermodynamic
properties and phase structures of strong 't Hooft coupling
conformal field theories by studying the thermodynamics of AAdS
black holes. An interesting application of the AdS/CFT
correspondence is the interpretation of Hawking-Page phase
transition between thermal AdS and AAdS black hole as the
confinement-deconfinement phases of the Yang-Mills (dual gauge)
theory defined on the AdS boundary \cite{Wit}.

This conjecture is now a fundamental concept that furnishes a
means for calculating the action and thermodynamic quantities
intrinsically without reliance on any reference spacetime
\cite{Sken1,BK,Od1}. It has
extended to the case of asymptotically de Sitter spacetimes \cite{Stro,Deh1}%
. Although the (A)dS/CFT correspondence applies for the case of a
specially infinite boundary, it was also employed for the
computation of the conserved and thermodynamic quantities in the
case of a finite boundary \cite{Deh2}. This conjecture has also
been applied for the case of black objects with constant negative
or zero curvature horizons \cite{Deh3,Deh4}.

For AAdS spacetimes, the presence of a negative cosmological
constant makes it possible to have a large variety of black
holes/branes, whose event horizons are hypersurfaces with
positive, negative, or zero scalar curvatures \cite{Man1}. The
AAdS rotating solution of Einstein's equation with cylindrical and
toroidal horizon and its extension to include the electromagnetic
field have been considered in Ref. \cite{Lem1}. The generalization
of this AAdS charged rotating solution of Einstein-Maxwell's
equation to the higher dimensions has been done in Ref.
\cite{Awad2}. Many authors have been considered thermodynamics and
stability conditions of these black holes \cite{Pec,Deh3}. In this
paper, we study the phase behavior of the charged rotating black
branes in $(n+1)$ dimensions with zero curvature horizon and show
that there is no Hawking-Page phase transition in spite of the
angular momentum of the branes. This is in commensurable with the
fact that there is no Hawking-Page phase transition for black
object whose horizon is diffeomorphic to $\mathbb{R}^p$ ($p$-brane
solution) and therefore the system is always in the high
temperature phase \cite{Wit}. According to the AdS/CFT dictionary,
this means that the corresponding field theory, which now lives on
$S^1\times \mathbb{R}^p$, has no phase transition as a function of
temperature in the large $\mathcal{N}$ limit.

The outline of our paper is as follows. We review the basic
formalism in Sec. \ref{Gen}. In Sec. \ref{Metr} we consider the
$(n+1)$-dimensional AAdS charged rotating black brane. We also
compute the Euclidean actions of the system both in the canonical
and the grand-canonical ensemble, and obtain the logarithmic
divergences associated to the Weyl anomalies and matter fields. In
Sec. \ref{Therm}, we study the thermodynamics of the brane, and
perform a thermal stability analysis. Also, the logarithmic
correction to the Bekenstein-Hawking entropy of the black brane is
obtained. We finish our paper with some concluding remarks.

\section{General Formalism \label{Gen}}

The gravitational action for Einstein-Maxwell theory in $(n+1)$ dimensions
for AAdS spacetimes is
\begin{equation}
I_G=-\frac 1{16\pi }\int_{\mathcal{M}}d^{n+1}x\sqrt{_{-}g}\left(
R-2\Lambda
-F^{\mu \nu }F_{\mu \nu }\right) +\frac 1{8\pi }\int_{\partial \mathcal{M}%
}d^nx\sqrt{_{-}\gamma }K(\gamma ),  \label{Actg}
\end{equation}
where $F_{\mu \nu }=\partial _\mu A_\nu -\partial _\nu A_\mu $ is
the electromagnetic tensor field and$\ A_\mu $ is the vector
potential. The first term is the Einstein-Hilbert volume term with
negative cosmological constant $\Lambda =-n(n-1)/2l^2$ and the
second term is the Gibbons Hawking boundary term which is chosen
such that the variational principle is well defined. The manifold
$\mathcal{M}$ has metric $g_{\mu \nu }$ and covariant derivative
$\nabla _\mu $. $K$ is the trace of the extrinsic curvature
$K^{\mu \nu }$ of any boundary(ies) $\partial \mathcal{M}$ of the
manifold $\mathcal{M}$, with induced metric(s) $\gamma _{i,j}$.
The AdS/CFT
correspondence states that if the metric near the conformal boundary ($%
x\rightarrow 0$) can be expanded in the AAdS form
\begin{equation}
ds^2=\frac{dx^2}{l^2x^2}+\frac 1{x^2}h _{ij}dx^idx^j,
\label{metc}
\end{equation}
with nondegenerate metric $h_{ij} $, then one may remove the non
logarithmic divergent terms in the action by adding a counterterm
action $I_{ct}$ which is a functional of the boundary curvature
invariants. It is worthwhile to mention that the induced metric on
the boundary is $\gamma_{ij}=h_{ij}/\epsilon^2$ where $\epsilon\ll
1$. The counterterm for asymptotically AdS spacetimes up to seven
dimensions is
\begin{eqnarray}
I_{ct} &=&\frac{1}{8\pi }\int_{\partial \mathcal{M}_\infty }d^nx\sqrt{%
-\gamma }\{\frac{n-1}l-\frac{l\Theta (n-3)}{2(n-2)}R  \nonumber \\
&&-\frac{l^3\Theta (n-5)}{2(n-4)(n-2)^2}\left( R_{ab}R^{ab}-\frac
n{4(n-1)}R^2\right) +...\}, \label{Actct1}
\end{eqnarray}
where $R$, $R_{abcd}$, and $R_{ab}$ are the Ricci scalar, Riemann,
and Ricci tensors of the boundary metric $\gamma _{ab}$ and
$\Theta (x)$ is the step function which is equal to one for $x\geq
0$ and zero otherwise. Although other counterterms (of higher mass
dimension) may be added to $I_{ct}$, they will make no
contribution to the evaluation of the action or Hamiltonian due to
the rate at which they decrease toward infinity, and we shall not
consider them in our analysis here. These counterterms have been
used by many authors for a wide variety of the spacetimes,
including Schwarzschild-AdS, topological Schwarzschild-AdS,
Kerr-AdS, Taub-NUT-AdS, Taub-bolt-AdS, and Taub-bolt-Kerr-AdS
\cite{EJM}.

Of course, for even $n$ one has logarithmic divergences in the partition
function which can be related to the Weyl anomalies in the dual conformal
field theory \cite{Sken1}. These logarithmic divergences associated with the
Weyl anomalies of the dual field theory for $n=4$ and $n=6$ are \cite{Sken2}%

\begin{eqnarray}
I_{\mathrm{\log }} &=&-\frac{\ln \epsilon }{64\pi l^3}\int
d^4x\sqrt{-h
^0}\left[ (R_{ij}^0)R^{(0)ij}-\frac 13(R^0)^2\right] ,  \label{Ano4} \\
I_{\log } &=&\frac{\ln \epsilon }{8^4\pi l^3}\int d^6x\sqrt{-h ^0}%
\{\frac 3{50}(R^0)^3+R^{(0)ij}R^{(0)kl}R_{ijkl}^0 -\frac 12R^0R^{(0)ij}R_{ij}^0 \nonumber \\
&&\hspace{3cm} +\frac 15R^{(0)ij}D_iD_jR^0-\frac 12R^{(0)ij}\Box
^0R_{ij}^0+\frac 1{20}R^0\Box ^0R^0\}.  \label{Ano6}
\end{eqnarray}
In Eqs. (\ref{Ano4}) and (\ref{Ano6}) $R^0$ and $R^{(0)ij}$ are
the Ricci scalar and Ricci tensor of the leading order metric $h
^0$ in the following expansion:
\begin{equation}
h _{ij}=h _{ij}^0+x^2h _{ij}^2+x^4h _{ij}^4+... , \label{exp}
\end{equation}
and $D_i$ is the covariant derivative constructed by the leading
order metric $h ^0$. Also, one should note that the inclusion of
matter fields in the gravitational action produces an additional
logarithmic divergence in the action for even $n$. This
logarithmic divergence for $n=4$ and $n=6$ are \cite{Robin}
\begin{eqnarray}
I_{\log }^{\mathrm{em}} &=&\frac{\ln \epsilon }{64\pi l}\int d^4x\sqrt{%
h ^0}F^{(0)ij}F_{ij}^0, \label{logem4} \\
I_{\log }^{\mathrm{em}} &=&\frac{\ln \epsilon }{8\pi l^3}\int d^6x\sqrt{%
-h ^0}\{ \frac 1{16}R^0F^{(0)ij}F_{ij}^0-\frac
18R^{(0)ij}F_i^{(0)l}F_{jl}^0  \nonumber \\
&&\hspace{3.5cm} +\frac
1{64}F^{(0)ij}(D_jD^kF_{ki}^0-D_iD^kF_{kj}^0)\}. \label{logem6}
\end{eqnarray}
where $F_{ij}^0$ is the leading term of electromagnetic field on
the conformal boundary. All the contractions in Eqs.
(\ref{logem4}) and (\ref {logem6}) should be done by the leading
order metric $h ^0$. In $4<n<7$ the matter field will cause a
power law divergence in the action which can be removed by a
counterterm of the form \cite{Robin}
\begin{equation}
I_{ct}^{\mathrm{em}}=\frac 1{256\pi }\int d^nx\sqrt{_{-}\gamma }\frac{(n-8)}{%
(n-4)}\Theta (n-5)F^{ij}F_{ij}.  \label{Actct2}
\end{equation}

Thus, the total action can be written as a linear combination of
the gravity term (\ref{Actg}), the logarithmic divergences
(\ref{Logc1}), (\ref{Logc2}), (\ref{logem4}), and (\ref{logem6})
and the counterterms (\ref{Actct1}) and (\ref{Actct2}). We will
show that for the charged rotating black branes investigated in
this paper, all the logarithmic divergences and the counterterm
$I_{\mathrm{ct}}^{\mathrm{em}}$ are zero and therefore the total
renormalized action is
\begin{equation}
I=I_G+I_{\mathrm{ct}}. \label{Acttot}
\end{equation}
In order to obtain the Einstein-Maxwell equations by the variation
of the volume integral with respect to the fields, one should
impose the boundary condition $\delta A_\mu =0$ on $\partial
\mathcal{M}$. Thus the action (\ref {Acttot}) is appropriate to
study the grand-canonical ensemble with fixed electric potential
\cite{Cal}. To study the canonical ensemble with fixed electric
charge one should impose the boundary condition $\delta
(n^aF_{ab})=0$, and therefore the total action is \cite{Haw1}
\begin{equation}
\stackrel{\sim }{I}=I-\frac 1{4\pi }\int_{\partial \mathcal{M}_\infty }d^nx%
\sqrt{_{-}\gamma }n_aF^{ab}A_b.  \label{Actcan}
\end{equation}
Having the total finite action, one can use the Brown and York definition
\cite{Brown} to construct a divergence free stress-energy tensor as
\begin{eqnarray}
T^{ab} &=&\frac 1{8\pi }\{(K^{ab}-K\gamma ^{ab})-\frac{n-1}l\gamma
^{ab}+\frac l{n-2}(R^{ab}-\frac 12R\gamma ^{ab})  \nonumber \\
&&\ +\frac{l^3\Theta (n-5)}{(n-4)(n-2)^2}[-\frac 12\gamma
^{ab}(R^{cd}R_{cd}-\frac n{4(n-1)}R^2)-\frac n{(2n-2)}RR^{ab}  \nonumber \\
&&\ +2R_{cd}R^{acbd}-\frac{n-2}{2(n-1)}\nabla ^a\nabla ^bR+\nabla
^2R^{ab}-\frac 1{2(n-1)}\gamma ^{ab}\nabla ^2R]+...\}.  \label{Stres}
\end{eqnarray}
The above stress-tensor is divergence free for $n\leq 6$, but we
can always add more counterterms to have a finite action in higher
dimensions (see e.g. Ref. \cite{Krau}).

The conserved charges associated to a Killing vector $\xi ^a$ is
\begin{equation}
\mathcal{Q}(\xi )=\int_{\mathcal{B}}d^nx\sqrt{\sigma }n^aT_{ab}\xi ^b,
\label{Con}
\end{equation}
where $\sigma $ is the determinant of the metric $\sigma _{ij}$
and $N$ is the lapse function, appearing in the ADM-like
decomposition of the boundary metric
\begin{equation}
ds^2=-N^2dt^2+\sigma _{ab}(dx^a+N^adt)(dx^b+N^bdt).
\end{equation}
For boundaries with timelike Killing vector ($\xi =\partial /\partial t$)
and rotational Killing vector field $(\zeta =\partial /\partial \phi )$ one
obtains the conserved mass and angular momentum of the system enclosed by
the boundary $\mathcal{B}$. In the context of AdS/CFT correspondence, the
limit in which the boundary $\mathcal{B}$ becomes infinite $(\mathcal{B}%
_\infty )$ is taken, and the counterterm prescription ensures that the
action and conserved charges are finite. No embedding of the surface $%
\mathcal{B}$ in to a reference of spacetime is required and the quantities
which are computed are intrinsic to the spacetimes.

\section{The Action and Thermodynamic Quantities of AAdS Charged Rotating
Black Brane\label{Metr}}

The metric of $(n+1)$-dimensional AAdS charged rotating black brane with $k$
rotation parameters is \cite{Awad2}
\begin{eqnarray}
ds^2 &=&-f(r)\left( \Xi dt-{{\sum_{i=1}^k}}a_id\phi _i\right) ^2+\frac{r^2}{%
l^4}{{\sum_{i=1}^k}}\left( a_idt-\Xi l^2d\phi _i\right) ^2  \nonumber \\
&&\ \text{ }+\frac{dr^2}{f(r)}-\frac{r^2}{l^2}{\sum_{i<j}^k}(a_id\phi
_j-a_jd\phi _i)^2+r^2d\Omega ^2,  \label{met2}
\end{eqnarray}
where $\Xi =\sqrt{1+\sum_i^ka_i^2/l^2}$ and $d\Omega ^2$ is the Euclidean
metric on the $\left( n-1-k\right) $-dimensional submanifold. The maximum
number of rotation parameters in $(n+1)$ dimensions is $[(n+1)/2]$, where $%
[x]$ denotes the integer part of $x$. In Eq. (\ref{met2}) $f(r)$ is
\begin{equation}
f(r)=\frac{r^2}{l^2}-\frac m{r^{n-2}}+\frac{q^2}{r^{2n-4}},  \label{Fg}
\end{equation}
and the gauge potential is given by
\begin{equation}
A_\mu =-\sqrt{\frac{n-1}{2n-4}}\frac q{r^{n-2}}\left( \Xi \delta
_\mu ^0-\delta _\mu ^ia_i\right),\hspace{1.0cm}(\text{no sum on
}i). \label{Vecp}
\end{equation}
As in the case of rotating black hole solutions of Einstein's gravity, the
above metric given by Eqs. (\ref{met2})-(\ref{Vecp}) has two types of
Killing and event horizons. The Killing horizon is a null surface whose null
generators are tangent to a Killing field. It was proved that a stationary
black hole event horizon should be a Killing horizon in the four-dimensional
Einstein gravity \cite{Haw1}. This fact is also true for this $(n+1)$%
-dimensional metric and the Killing vector
\begin{equation}
\chi =\partial _t+{{{\sum_{i=1}^k}}}\Omega _i\partial _{\phi _i},
\label{Kil}
\end{equation}
is the null generator of the event horizon. The metric of Eqs. (\ref{met2})-(%
\ref{Vecp}) has two inner and outer event horizons located at $r_{-}$ and $%
r_{+}$, if the metric parameters $m$ and $q$ are chosen to be suitable \cite{Awad2}%
. For later use in the thermodynamics of the black brane, it is
better to present an expression for the critical value of the
charge in term of the radius of the event horizon $r_+$. It is
easy to show that the metric has two inner and outer horizons
provided the charge parameter, $q$ is less than
$q_{\mathrm{crit}}$ given as
\begin{equation}
q_{\mathrm{crit}}=\sqrt{\frac{n}{n-2}} \frac{r_+^{n-1}}{l}.
\label{qcrit}
\end{equation}
In the case that $q=q_{\mathrm{crit}}$, we will have an extreme black brane.

The mass, the angular momenta, the Hawking temperature and the angular
velocities of the outer event horizon have been calculated in Ref. \cite
{Awad2}. We bring them here for later use
\begin{eqnarray}
M &=&\frac{V_{n-1}}{16\pi }m\left[ n\Xi ^2-1\right],  \label{Mass} \\
J_i &=&\frac{V_{n-1}}{16\pi }n\Xi ma_i ,  \label{Angmom} \\
T &=&\frac 1{\beta _{+}}=\frac{nr_{+}^{(2n-2)}-(n-2)q^2l^2}{4\pi l^2\Xi
r_{+}^{(2n-3)}},  \label{Temp} \\
\Omega _j &=&\frac{a_j}{\Xi l^2},  \label{Om}
\end{eqnarray}
where $V_{n-1}$ denotes the volume of the hypersurface boundary
$\mathcal{B}$ at constant $t$ and $r$, and $\beta _{+}$ is the
inverse Hawking
temperature. Equation (\ref{qcrit}) shows that the temperature $T$ in Eq. (%
\ref{Temp}) is positive for the allowed values of the metric parameters and
vanishes for the extremal solution.

To obtain the total action we first calculate the logarithmic
divergences due to the Weyl anomaly and matter field given in Eqs.
(\ref{Ano4}), (\ref{Ano6}), (\ref {logem4}), and (\ref{logem6}).
The leading metric $h _{ij}^0$ can be obtained as
\begin{equation}
h _{ij}^0dx^idx^j=-\frac 1{l^2}dt^2+d\phi ^2+d\Omega ^2.
\label{gamma0}
\end{equation}
Therefore the curvature scalar $R^0(h ^0)$ and Ricci tensor $%
R_{ij}^0(h ^0)$ are zero. Also it is easy to show that $F_{ij}^0$
in Eqs. (\ref{logem4}) and (\ref{logem6}) vanishes. Thus, all the
logarithmic divergences for the $(n+1)$-dimensional charged
rotating black brane are zero. It is also a matter of calculation
to show that the counterterm action due to the
electromagnetic field in Eq. (\ref{Actct2}) is zero. Thus, using Eqs. (\ref{Actg}%
), (\ref{Actct1}), (\ref{Acttot}), and (\ref{Actcan}), the
Euclidean actions in the grand-canonical and the canonical
ensemble can be calculated as
\begin{eqnarray}
I &=&-\frac{\beta _{+}V_{n-1}}{16\pi }\frac{r_{+}^{(2n-2)}+q^2l^2}{%
r_{+}^{(n-2)}l^2},  \label{Igcm} \\
\stackrel{\sim }{I} &=&-\frac{\beta _{+}V_{n-1}}{16\pi }\frac{%
r_{+}^{(2n-2)}-(2n-3)q^2l^2}{r_{+}^{(n-2)}l^2}.  \label{Icm}
\end{eqnarray}
The electric charge $Q$, can be found by calculating the flux of the
electromagnetic field at infinity, yielding
\begin{equation}
Q=\frac{\Xi V_{n-1}}{4\pi }\sqrt{\frac{(n-1)(n-2)}2}q.  \label{Charg}
\end{equation}
The electric potential $\Phi $, measured at infinity with respect to the
horizon, is defined by \cite{Cal}
\[
\Phi =A_\mu \chi ^\mu \left| _{r\rightarrow \infty }-A_\mu \chi ^\mu \right|
_{r=r_{+}},
\]
where $\chi $ is the null generators of the event horizon given by Eq. (\ref
{Kil}). One obtains
\begin{equation}
\Phi =\sqrt{\frac{(n-1)}{2(n-2)}}\frac q{\Xi r_{+}^{(n-2)}}.
\label{Pot}
\end{equation}
Since the area law of the entropy is universal, and applies to all kinds of
black holes/branes \cite{Beck}, the entropy is
\begin{equation}
S=\frac{\Xi V_{n-1}}4r_{+}^{(n-1)}.  \label{Entropy}
\end{equation}
For $n=3$, these quantities given in Eqs.
(\ref{Mass})-(\ref{Entropy}) reduce to those calculated in Ref.
\cite{Deh3}.

\section{Thermodynamics of black brane\label{Therm}}

\subsection{Energy as a function of entropy, angular momentum, and charge}

We first obtain the mass as a function of the extensive quantities
$S$, $J$, and $Q$. Using the expression for the entropy, the mass,
the angular momenta, and the charge given in Eqs. (\ref{Mass}),
(\ref{Angmom}), (\ref {Charg}), (\ref{Entropy}), and the fact that
$f(r_{+})=0$, one can obtain a Smarr-type formula as
\begin{equation}
M(S,J,Q)=\frac{(nZ-1) \sqrt{\sum_i^k J_i^2}}{nl\sqrt{Z(Z-1)}},
\label{Smar}
\end{equation}
where $Z=\Xi ^2$ is the positive real root of the following
equation:

\begin{equation}
\left( Z-1\right) ^{(d-2)}-\frac Z{16S^2}\left\{ \frac{4\pi (n-1)(n-2)lSJ}{%
n[(n-1)(n-2)S^2+2\pi ^2Q^2l^2]}\right\} ^{(2n-2)}=0. \label{Zsmar}
\end{equation}
One may then regard the parameters $S$, $J$, and $Q$ as a complete
set of extensive parameters for the mass $M(S,J,Q)$ and define the
intensive parameters conjugate to $S$, $J$ and $Q$. These
quantities are the temperature, the angular velocities, and the
electric potential
\begin{equation}
T=\left( \frac{\partial M}{\partial S}\right) _{J,Q},\ \ \Omega
_i=\left( \frac{\partial M}{\partial J_i}\right) _{S,Q},\ \ \Phi
=\left( \frac{\partial M}{\partial Q}\right) _{S,J}. \label{Dsmar}
\end{equation}
It is a matter of straightforward calculation to show that the intensive
quantities calculated by Eq. (\ref{Dsmar}) coincide with Eqs. (\ref{Temp}), (%
\ref{Om}), and (\ref{Pot}) found in Sec. (\ref{Metr}). Thus, the
thermodynamic quantities calculated in Sec. (\ref{Metr}) satisfy
the first law of thermodynamics

\begin{equation}
dM=TdS+{{{\sum_{i=1}^k}}}\Omega _idJ_i+\Phi dQ.  \label{Flth}
\end{equation}

\subsection{Thermodynamic potentials}

We now obtain the thermodynamic potential in the grand-canonical and
canonical ensembles. Using the definition of the Gibbs potential $G(T,\Omega
,\Phi )=I/\beta $ , we obtain

\begin{equation}
G=-\frac{V_{n-1}}{16\pi }\left( \frac 2{n^2(n-1)(1-\sum_i^kl^2\Omega
_i^2)}\right) ^{n/2}\left( \gamma ^2+n^2(n-2)\Phi ^2\right) (\gamma
l)^{(n-2)},  \label{Gibbs}
\end{equation}
where
\begin{equation}
\gamma = \sqrt{2n-2}T\pi l+\sqrt{2(n-1)\pi ^2T^2l^2+n(n-2)^2\Phi ^2}.
\label{Ggibbs}
\end{equation}
Using the expressions (\ref{Temp}), (\ref{Om}), and (\ref{Pot})
for the inverse Hawking temperature, the angular velocities and
the electric potential, one obtains

\begin{equation}
G(T,\Omega ,\Phi )=M-TS-\sum_i^k\Omega _iJ_i-\Phi Q ,  \label{Lgibbs}
\end{equation}
which means that $G(T,\Omega ,\Phi )$ is, indeed, the Legendre
transformation of the $M(S,J_i,Q)$ with respect to $S,$ $J_i$, and
$Q$. It is a matter of straightforward calculation to show that
the extensive quantities

\begin{equation}
J_i=-\left( \frac{\partial G}{\partial \Omega _i}\right) _{T,\Phi },\ \
Q=-\left( \frac{\partial G}{\partial \Phi }\right) _{T,\Omega }, \ \
S=-\left( \frac{\partial G}{\partial T}\right) _{\Omega ,\Phi },
\label{Dgibbs}
\end{equation}
turn out to coincide precisely with the expressions
(\ref{Angmom}), (\ref {Charg}), and (\ref{Entropy}).

For the canonical ensemble, the Helmholtz free energy $F(T,J,Q)$ is defined
as

\begin{equation}
F(T,J,Q)=\frac{\stackrel{\sim }{I}}\beta +\sum_i^k\Omega _iJ_i,
\label{Helm}
\end{equation}
where $\stackrel{\sim }{I}$ is given by Eq. (\ref{Icm}). One can verify that
the conjugate quantities

\begin{equation}
\Omega _i=\left( \frac{\partial F}{\partial J_i}\right) _{T,Q}, \
\ \Phi
=\left( \frac{\partial F}{\partial Q}\right) _{T,J}, \ \ S=-\left( \frac{%
\partial F}{\partial T}\right) _{J,Q},  \label{Dhelm}
\end{equation}
agree with expressions (\ref{Om}), (\ref{Pot}), and
(\ref{Entropy}). Also it is worthwhile to mention that $F(T,J,Q)$
is the Legendre transformation of the $M(S,J_i,Q)$ with respect to
$S$, i.e

\begin{equation}
F(T,J,Q)=M-TS.  \label{Lhelm}
\end{equation}

\subsection{Stability in the canonical and the grand-canonical ensemble}

\begin{figure}[tbp]
\epsfxsize=10cm \centerline{\epsffile{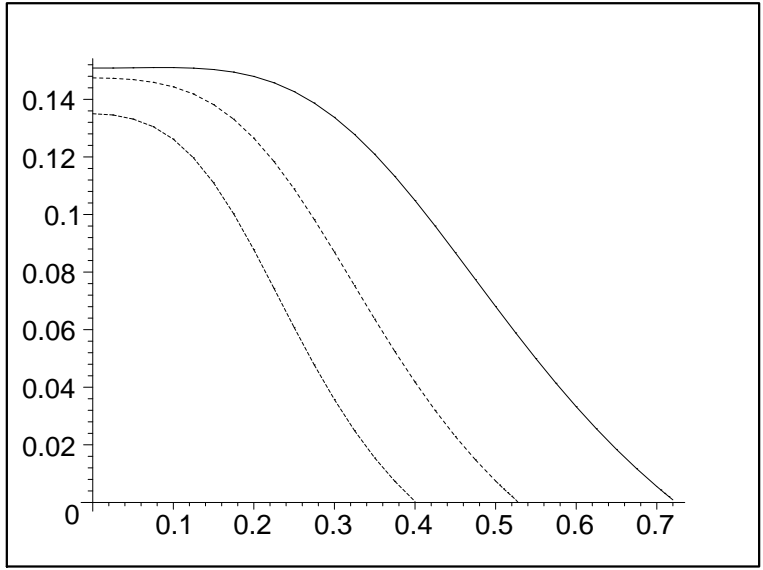}}
\caption{$C_{J,Q}$ versus $q$ for $l=1$, $r_+=0.8$, $n=4$ (solid),
$n=5$ (dotted), and $n=6$ (dashed).} \label{Figure1}
\end{figure}

The stability of a thermodynamic system with respect to the small variations
of the thermodynamic coordinates, is usually performed by analyzing the
behavior of the entropy $S(M,J,Q)$ around the equilibrium. The local
stability in any ensemble requires that $S(M,J,Q)$ be a convex function of
their extensive variables or its Legendre transformation must be a concave
function of their intensive variables. Thus, the local stability can in
principle be carried out by finding the determinant of the Hessian matrix of
$S$ with respect to its extensive variables $X_i$, $\mathbf{H}%
_{X_iX_j}^S=[\partial ^2S/\partial X_i\partial X_j]$, or the determinant of
the Hessian of the Gibbs function with respect to its intensive variables $%
Y_i$, $\mathbf{H}_{Y_iY_j}^G=[\partial ^2G/\partial Y_i\partial
Y_j]$ \cite {Cvet,Cal}. Also, one can perform the stability
analysis through the use of the Hessian matrix of the mass with
respect to its extensive parameters \cite {Gub}. In our case the
entropy $S$ is a function of the mass, angular momenta, the
charge. The number of thermodynamic variables depends on the
ensemble which is used. In the canonical ensemble, the charge and
the angular momenta are fixed parameters, and therefore the
positivity of the thermal capacity $C_{J,Q}=T(\partial S/\partial
T)_{J,Q}$ is sufficient to assure the local stability. The thermal
capacity $C_{J,Q}$ at constant charge and angular momenta is

\begin{eqnarray}
C_{J,Q} &=&\frac{\Xi V_{n-1}}%
4r^{(n-1)}[nr^{(2n-2)}-(n-2)q^2l^2][r^{(2n-2)}+q^2l^2]  \nonumber \\
&&\text{ }\times \ [(n-2)\Xi ^2+1]\{(n-2)q^4l^4[(3n-6)\Xi ^2-(n-3)]
\nonumber \\
&&\ -2q^2l^2r^{(2n-2)}[(3n-6)\Xi ^2-n^2+3]+nr^{(4n-4)}[(n+2)\Xi
^2-(n+1)]\}^{-1}. \label{Cap}
\end{eqnarray}
Figure \ref{Figure1} shows the behavior of the heat capacity as a
function of the charge parameter. It shows that $C_{J,Q}$ is
positive in various dimensions and goes to zero as $q$ approaches
its critical value (extreme black brane). Thus, the
$(n+1)$-dimensional AAdS charged rotating black brane is locally
stable in the canonical ensemble.

In the grand-canonical ensemble, we find it more convenient to
work with the Gibbs potential $G(T,\Omega _i,\Phi )$. Here the
thermodynamic variables are the temperature, the angular
velocities, and the electric potential. After some algebraic
manipulation, we obtain

\begin{equation}
\left| \mathbf{H}_{T,\Omega _i,\Phi }^G\right| =\left( \frac{n-3}n\right)
\left[ \frac{nm}{16\pi }\right] ^k\left[ \frac{(n-2)\Xi ^2+1}{r_{+}^n+(\frac{%
n-2}2)ml^2}\right] (l\Xi )^{2k+2}\Xi ^2r^{3n-4}. \label{dHes}
\end{equation}
As one can see from Eq. (\ref{dHes}), $\left| \mathbf{H}_{T,\Omega _i,\Phi
}^G\right| $ is positive for all the phase space, and therefore the $(n+1)$%
-dimensional AAdS charged rotating black brane is locally stable in the
grand-canonical ensemble. As mentioned in the previous paragraph, one can
also perform the local stability analysis through the use of the determinant
of the Hessian matrix of $M$ with respect to $S $, $J$ and $Q$, which has
the same result, since $\left| \mathbf{H}_{S,J_i,Q}^M\right| =\left| \mathbf{%
H}_{T,\Omega _i,\Phi }^G\right| ^{-1}$.

\subsection{ Logarithmic correction to the Bekenstein-Hawking entropy}

In recent years, there are several works in literature suggesting that for a
large class of black holes, the area law of the entropy receives additive
logarithmic corrections due to thermal fluctuation of the object around its
equilibrium \cite{Maj}. Typically, the corrected formula has the form
\begin{equation}
S=S_0-K\ln (S_0)+...,  \label{Logc1}
\end{equation}
where $S_0$ is the standard Bekenstein-Hawking term and $K$ is a
number. In Ref. \cite{Das} an expression has been found for the
leading-order correction of a generic thermodynamic system in
terms of the heat capacity $C$ as \cite {Das}
\begin{equation}
S=S_0-K\ln (CT^2).  \label{Logc2}
\end{equation}

Equation (\ref{Logc2}) has been considered by many authors for
Schwarzschild-AdS, Reissner-Nordstrom-AdS, BTZ, and slowly Kerr-AdS
spacetimes \cite{Od2}. Thus, it is worthwhile to investigate its application
for the charged rotating black brane considered in this paper. Using Eqs. (%
\ref{Temp}), (\ref{Entropy}), and (\ref{Cap}) with $q=0$, one
obtains

\begin{equation}
S=S_0-\frac{n+1}{2(n-1)}\ln (S_0)-\Gamma _n(\Xi ),  \label{Logq0}
\end{equation}
where $\Gamma _n(\Xi )$ is a positive constant depending on $\Xi $ and $n$.
Equation (\ref{Logq0}) shows that the correction of the entropy is
proportional to the logarithm of the area of the horizon. For small values
of $q$ the logarithmic correction in Eq. (\ref{Logc2}) can be expanded in
terms of the power of $S_0$ as
\begin{equation}
S=S_0-\frac{n+1}{2(n-1)}\ln (S_0)-\Gamma _n(\Xi )+\frac{l^2\Xi
^2[(n-4)\Xi ^2+2]}{16[(n+2)\Xi ^2-n-1]}\frac{q^2}{S_0^2}+...
\label{Logq}
\end{equation}
Again the leading term is a logarithmic term of the area.

\section{CLOSING REMARKS}

In this paper, we calculated the conserved quantities and the
Euclidean actions of the charged rotating black branes both in the
canonical and the grand-canonical ensemble through the use of
counterterms renormalization procedure. Also we obtained the
charge and electric potentials of the black brane in an arbitrary
dimension. We found that the logarithmic divergencies associated
with the Weyl anomalies and matter field are zero. We obtained a
Smarr-type formula for the mass as a function of the extensive
parameters $S$, $J$\, and $Q$, calculated the temperature, angular
velocity, and electric potential, and showed that these quantities
satisfy the first law of thermodynamics. Using the conserved
quantities and the Euclidean actions, the thermodynamics
potentials of the system in the canonical and grand-canonical
ensemble were calculated. We found that the Helmholtz free energy
$F(T,J,Q)$ is a Legendre transformation of the mass with respect
to $S$ and the Gibbs potential is a Legendre transformation of the
mass with respect to $S,$\ $J$\, and $Q$\ in the grand-canonical
ensemble.

Also, we studied the phase behavior of the charged rotating black branes in $%
(n+1)$ dimensions and showed that there is no Hawking-Page phase
transition in spite of the angular momentum of the branes. Indeed,
we calculated the heat capacity and the determinant of the Hessian
matrix of the Gibbs potential with respect to $S$, $J$, and $Q$ of
the black brane and found that they are positive for all the phase
space, which means that the brane is stable for all the allowed
values of the metric parameters discussed in Sec. \ref{Metr}. This
analysis has also be done through the use of the determinant of
the Hessian matrix of $M(S,J,Q)$ with respect its extensive
variables and we got the same phase behavior. This phase behavior
is commensurate with the fact that there is no Hawking-Page
transition for a black object whose horizon is diffeomorphic to
$\mathbb{R}^p$ and therefore the system is always in the high
temperature phase \cite{Wit}.

Finally, we obtained the logarithmic correction of the entropy due
to the thermal fluctuation around the thermal equilibrium. For the
case of uncharged rotation black brane, we found that only a term
which is proportional to $\ln (\mathrm{area})$ will appear. But we
found that for the charged rotating black brane, the correction
contains other powers of the $area$ including the logarithmic
term. Verlinde drew a fundamental connection between the
holographic principle, the entropy formula for conformal field
theory, and the Friedman-Robertson-Walker equations for a closed
radiation dominated universe \cite{Ver}. Therefore the
investigation of the effect of the logarithmic corrections to the
Cardy-Verlinde formula when thermal fluctuations of the AAdS black
brane are taken into account and their influence on the braneworld
cosmology remains a subject for future.

\end{document}